\documentclass[aps,pra,10pt,superscriptaddress,twocolumn,hidelinks,longbibliography,floatfix]{revtex4-2}
\usepackage{amsmath}
\usepackage{graphicx}
\usepackage{hyperref}
\usepackage{blindtext}
\usepackage{amsfonts}
\usepackage{orcidlink}

\begin{document}

\title{Quantum-Like Correlations from Local Hidden-Variable Theories Under Conservation Law}

\author{Alejandro J. Garza}
\email{alejandro.garza@stjude.org}
\affiliation{Department of Physics, Central Michigan University, 
1200 S. Franklin St.
Mount Pleasant, Michigan 48859}
\altaffiliation{Current Affiliation: Department of Structural Biology, St. Jude 
Children's Research Hospital, 262 Danny Thomas Pl, Memphis, TN 38105}

\author{Jonte R. Hance\,\orcidlink{0000-0001-8587-7618}}
\email{jonte.hance@newcastle.ac.uk}
\affiliation{School of Computing, Newcastle University, 1 Science Square, Newcastle upon Tyne, NE4 5TG, UK}
\affiliation{Quantum Engineering Technology Laboratories, Department of Electrical and Electronic Engineering, University of Bristol, Woodland Road, Bristol, BS8 1US, UK}

\begin{abstract}
 The precision with which we can measure operators that do not commute with conserved quantities is limited by the need to preserve the associated global symmetries. We show how to construct a local hidden-variable model that violates Bell inequalities by interpreting this measurement error as altering the measure space of the hidden variables. This provides a physically-motivated example of a supermeasured model, where the statistical independence assumption used to form Bell inequalities is violated without there being a causal dependence of the measurement settings on the hidden variables (superdeterminism) or a causal dependence of the hidden variables on the measurement settings (retrocausality). The model also gives subtly different predictions to quantum mechanics, that could be tested experimentally. 
\end{abstract}
\maketitle

\emph{Introduction---}Bell's theorem states that quantum mechanics is incompatible with the notion of local realism, formalized as local hidden variable theories which obey statistical independence~\cite{Bell1964}. Since its introduction in 1964, multiple and increasingly sophisticated experiments have confirmed the predictions of quantum mechanics~\cite{Freedman1972,Aspect1981,Aspect1982,Weihs1998,Pan2000,Rowe2001}. Possible loopholes in early Bell tests \cite{larsson2014loopholes} such as the memory, locality, measurement independence, and detection loopholes have been closed in more recent years~\cite{Hensen2015,Giustina2015,Shalm2015,Rosenfeld2017}.
From this we can conclude that one of the assumptions used to craft local hidden variable theories must be violated by quantum mechanics (and the physical world). 

The detection loophole (also known as the fair-sampling loophole) is based on the fact that, if certain experimental outcomes are selectively discarded or lost, then local hidden-variable theories can reproduce the results of quantum mechanics~\cite{Pearle1970}. This loophole has been closed by experiments using detectors with efficiencies above the threshold required for a loophole-free Bell test~\cite{Rowe2001,Giustina2015,Hensen2015}. However, one could conjecture a model where, by violating Statistical Independence (the assumption of no correlation between measurement choices and hidden variables), even in these loophole-free Bell tests, certain outcomes are selectively discarded. Rather than this being because of a causal dependence of the measurement choices on the hidden variables (superdeterminism) \cite{Hossenfelder2020Rethinking}, or vice-versa (retrocausality) \cite{Wharton2019Reformulations,adlam2023taxonomy}, this could be because of a physically-motivated alteration to the measure space, such that certain combinations of measurement settings and hidden variables are counterfactually restricted (i.e., the model being supermeasured) \cite{Hance_2024}.

Here, we propose a model where certain outcomes cannot be observed, not because of detector efficiency, but 
because they would violate conservation laws. Using the Wigner-Araki-Yanase (WAY) theorem~\cite{Wigner1952,Araki1960,Marvian2012,Loveridge2011} and Ozawa's accuracy bound for quantum measurements~\cite{Ozawa2002}, we show how a local hidden-variables model could imitate quantum correlations under conservation laws, so long as it violates Statistical Independence. The model also predicts classical behavior at the limit of large system size, and subtle differences with respect to quantum mechanical predictions that could be observed experimentally (e.g., slight differentiation in correlations between Bell singlet and triplet states originating from their distinct angular momenta). 
 
\emph{Results---}Consider a Bell experiment starting from a singlet pair of spin $\frac{1}{2}$ qubits
\begin{equation}
  |\psi^-\rangle = \frac{|01\rangle - |10\rangle}{\sqrt{2}}, 
\end{equation}
where  $|ab\rangle \equiv |a\rangle \otimes |b\rangle$. 
Let $\sigma_\alpha$ denote a spin observable measured along a direction in the $xz$-plane at 
an angle $\alpha$ from the $z$-axis. Thus,
\begin{equation}
  \sigma_\alpha = \sin(\alpha) \sigma_x + \cos(\alpha) \sigma_z,
\end{equation}
where $\sigma_j$ is the Pauli-$j$ matrix. 
The experiment measures $\langle  \sigma_\alpha \otimes \sigma_\beta \rangle$, 
whose quantum mechanical expectation value depends only on
$\theta = |\alpha - \beta|$ and is given by
\begin{equation}
  E^{qm}(\theta) = \langle \psi^- |  \sigma_\alpha \otimes \sigma_\beta  | \psi^-  \rangle = - \cos (\theta).
  \label{eq:EQM}
\end{equation}
A response function for a (Bell-)local hidden variable theory can be expressed as~\cite{Bell1964,Hance2022}
\begin{equation}
  E^{hv}(\alpha,\beta) = \int_{\Lambda} S_1(\alpha,\lambda) S_2(\beta,\lambda) \rho(\lambda) \mu(d\lambda), 
  \label{eq:integral0}
\end{equation}
where $\lambda \in \Lambda$ is a single or a set of hidden variables, $\mu(d\lambda)$ a measure, $\rho(\lambda)$ the probability density function of $\lambda$, and $S_i(\alpha,\lambda) = \pm 1$ denotes the outcome of a measurement for the $i$th qubit. The probability density function $\rho(\lambda)$ is independent of $\alpha$ and $\beta$; only the outcomes $S_i(\alpha,\lambda)$ depend on the detector settings. Because of Bell-locality, along with the facts that $|S_i(\alpha,\lambda)| = 1$ and $\int_\Lambda \rho(\lambda) \mu(d\lambda) = 1$, local hidden-variable theories are constrained by Bell-type inequalities which are violated by quantum mechanics~\cite{Bell1964}.

If the measurement process introduced bias dependent on $\alpha$ and $\beta$ that precluded outcomes from certain values of $\lambda$ from being observed, then the measure space $(\Lambda, \mathcal{A}, \mu)$ would be altered by the experiment settings and Bell-type inequalities could be violated (as occurs in the detection loophole). This is the case despite the fact that the underlying $\rho(\lambda)$ remains independent of the measurement settings. This possibility is referred to as the hidden-variable theory being Supermeasured~\cite{Hance2022}.
In the presence of symmetries, measurements do have inherent errors imposed by the WAY theorem, which states that observables that fail to commute with a conserved quantity cannot be measured exactly~\cite{Wigner1952,Araki1960,Marvian2012,Loveridge2011}.
We here demonstrate that, for spin qubits, if the error predicted by the WAY theorem to preserve 
angular momentum during measurement is interpreted as a bias that changes the measure space $(\Lambda, \mathcal{A}, \mu)$, then this allows for such hidden-variables theory to reproduce the results of quantum mechanics.

For our base hidden-variables model (i.e., not accounting for WAY error), we employ the same model used by Bell~\cite{Bell1964} in his original paper: $\lambda$ is an angle associated with the direction of a spin vector on the measurement plane, $\rho(\lambda)$ is uniform, and a measurement taken at any particular angle gives the result $S_i(\alpha,\lambda) = +1$ if $\alpha$ is within $\pi/2$ of $\lambda$, and $S_i(\alpha,\lambda) = -1$  otherwise. This yields a piecewise linear response function
\begin{align}
  E^{hv}(\theta)  & = - \frac{1}{\pi} \int\limits_0^{\pi-\theta} d\lambda + \frac{1}{\pi} \int\limits_{\pi-\theta}^\pi d\lambda \label{eq:EHV} \\ 
  & = (2/\pi)\theta - 1, \nonumber
\end{align}
defined for $\theta \in [0,\pi]$, which gives the response profile over $ \theta \in [0,2\pi]$ from symmetry - i.e.,
\begin{equation}
    E^{hv}_{\theta \in [\pi,2\pi]} (\theta) = -E^{hv}_{\theta \in [0,\pi]} (\theta - \pi).
\end{equation}

Now we focus on the measurement process and quantify the WAY error that arises from the need to preserve angular momentum globally. Our model to describe a measurement is based on that used by Ozawa~\cite{Ozawa2002}: a measurement requires an interaction between the system $|\psi^-\rangle$ and a probe $|\xi\rangle$ of that is part of an apparatus or environment performing the measurement. The time evolution of the composite system $|\psi^-\rangle \otimes |\xi\rangle$ during the measurement process is then described by a unitary operator $U_{\alpha\beta}$. Therefore, 
\begin{equation}
  U_{\alpha\beta} |\psi^-\rangle \otimes |\xi\rangle = | \alpha^\pm \beta^\mp \rangle \otimes | \xi^{\pm}_{\mp} \rangle,
  \label{eq:evolution} 
\end{equation}
where $| \xi^{\pm}_{\mp} \rangle$ represents distinguishable states of the probe associated with the various measurement outcomes, and
\begin{eqnarray}
  |\alpha^+\rangle &= \cos(\alpha/2)|0\rangle + \sin(\alpha/2)|1\rangle,\\
  |\alpha^-\rangle &= -\sin(\alpha/2)|0\rangle + \cos(\alpha/2)|1\rangle. 
\end{eqnarray}
In general, taking $|\psi^-\rangle$ to its final state $| \alpha^\pm \beta^\mp \rangle$ does not preserve angular momentum. 
We assume that $U_{\alpha\beta}$ is a unitary that acts on both $|\psi^-\rangle$ and the smallest subsystem of the apparatus (or environment) capable of acting as a probe for which angular momentum is conserved across $|\psi^-\rangle \otimes |\xi\rangle$ during the measurement. Note that $|\xi\rangle$ may then depend on $\alpha$ and $\beta$.
We omit these dependencies to make the notation less cumbersome, but will recall them when relevant. 

Under this measurement model, $\langle \sigma_\alpha \otimes \sigma_\beta \rangle$ is determined indirectly: the unitary $U_{\alpha\beta}$ couples $|\psi^-\rangle$ and $|\xi\rangle$, and takes them to their final states (Eq.~\eqref{eq:evolution}). The correlation is then inferred based on whether the outcomes of measuring $\langle \sigma_\alpha \rangle $ for the first qubit and $\langle \sigma_\beta \rangle $ for the second one yield the same or opposite signs. The experiment therefore measures $ \langle U_{\alpha\beta}^\dag ( \sigma_\alpha \otimes \sigma_\alpha \otimes I) U_{\alpha\beta}\rangle$ for $|\psi^-\rangle \otimes |\xi\rangle$, which has the same expectation value and variance as $\sigma_\alpha \otimes \sigma_\beta$ for $|\psi^-\rangle$. That is, $U_{\alpha\beta}^\dag ( \sigma_\alpha \otimes \sigma_\alpha \otimes I) U_{\alpha\beta}$ is directly associated with what is known as the pointer observable of the probe~\cite{Loveridge2011,Ozawa2002} in its post-measurement state.

Based on this fact, and inspired by the works of Ozawa~\cite{Ozawa2002,Ozawa2003} and Busch~\cite{Busch2009} on the limits of quantum measurement, we define the following deviation operator
\begin{equation}
  D = U_{\alpha\beta}^\dag (\sigma_\alpha \otimes \sigma_\alpha \otimes I) U_{\alpha\beta} - \sigma_\alpha \otimes \sigma_\beta \otimes I, 
  \label{eq:deviationD}
\end{equation}
to estimate the measurement error with respect to the true expectation value of $\sigma_\alpha \otimes \sigma_\beta$. While formally $\langle D \rangle = 0$ for $|\psi^-\rangle \otimes |\xi\rangle$, conservation of angular momentum together with the uncertainty principle impose restrictions on $\langle  D^2 \rangle$. 
Note that $\langle  D^2 \rangle \geq (\Delta D)^2 = \langle D^2 \rangle - \langle D \rangle^2$, and let $L$ be an additive conserved quantity such as angular momentum. Then we have from the Robertson uncertainty relation that
\begin{equation}
  \langle  D^2 \rangle \geq \frac{1}{4}\frac{|\langle [D,L] \rangle|^2}{(\Delta L)^2 }. 
  \label{eq:robertson}
\end{equation}
Consider conservation of the $y$ component of total angular momentum in $|\psi^-\rangle \otimes |\xi\rangle$, i.e. 
$L = J_y = J_y^\psi + J_y^\xi$.
We can evaluate the numerator on the right hand side of Eq.~\eqref{eq:robertson} by invoking the conservation law $L = U_{\alpha\beta}^\dag L U_{\alpha\beta}$, which implies that 
\begin{equation}
\begin{split}
  [U_{\alpha\beta}^\dag ( \sigma_\alpha \otimes \sigma_\alpha \otimes I) U_{\alpha\beta},L] = \\
  U_{\alpha\beta}^\dag [( \sigma_\alpha \otimes \sigma_\alpha \otimes I),L] U_{\alpha\beta}. 
  \label{eq:acl0}
\end{split}
\end{equation}
Using Eqs.~\eqref{eq:evolution} and \eqref{eq:acl0}, we obtain for $|\psi^-\rangle \otimes |\xi\rangle$ with 
$L = J_y$ the following bound for the expected squared deviation
\begin{equation}
  \langle  D^2 \rangle \geq \frac{1}{4}\frac{|\sin\theta |^2}{(\Delta L)^2 },
  \label{eq:acl1}
\end{equation}
with $\theta = |\alpha - \beta|$. This inequality and its derivation is similar to that of Ozawa's trade-off bound for the WAY theorem~\cite{Loveridge2011,Ozawa2002} except that we focus the deviation introduced into the system instead of the noise on the probe of the apparatus. 

\begin{figure}[t]
  \centering
  \includegraphics[width=0.98\linewidth]{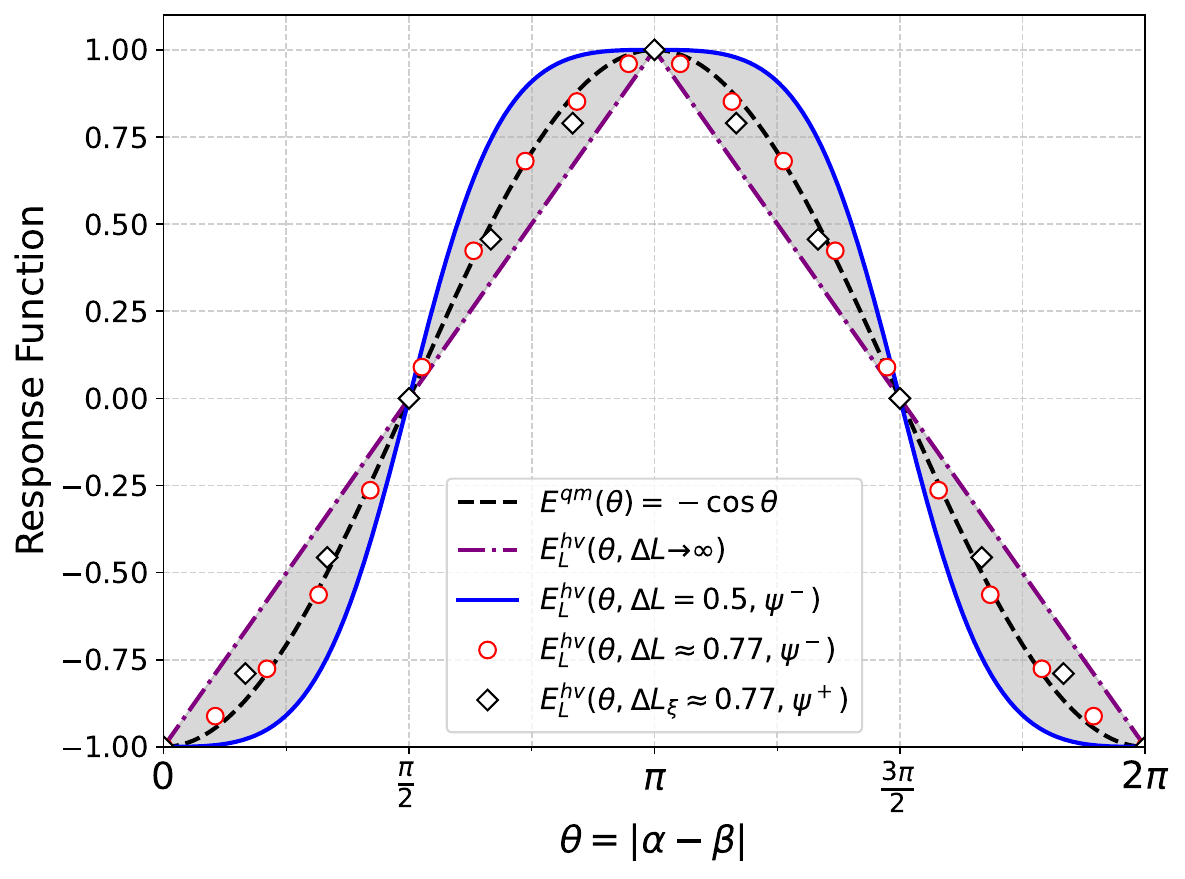}
  \caption{Bell test correlation response functions predicted by quantum mechanics ($E^{qm}$) and hidden-variable models considering conservation of angular momentum ($E_L^{hv}$) with various values of $\Delta L$ for the Bell singlet $|\psi^-\rangle$ and triplet $|\psi^+\rangle$.}
  \label{fig:fig1}
\end{figure}

Equation~\eqref{eq:acl0} indicates that, depending on $\theta$, introducing deviations in the system is necessary to preserve angular momentum during measurement. Thus, if there were hidden variables, the results associated with certain values of $\lambda$ will not be observed at certain values of $\theta$. Also, in a hidden-variable theory, such deviations must be deterministic.
Suppose then that $\lambda$ is shifted to satisfy Eq.~\eqref{eq:acl1} so that values that are within $b = \pm \langle D^2 \rangle^{1/2} = \pm \sin\theta/(2\Delta L)$ of $\theta$ are excluded from  $\Lambda$. The new normalization constant for the observed distribution is $\mathcal{N} = \pi - \sin\theta/(\Delta L)$, and the response function becomes
\begin{align}
  E^{hv}_L(\theta) & = \int\limits_0^{\pi - \theta - b} \frac{d\lambda}{\mathcal{N}} - \displaystyle\int\limits_{\pi - \theta + b}^{\pi} \frac{d\lambda}{\mathcal{N}} 
  \label{eq:ehvl}\\
  & = \frac{\Delta L (2\theta - \pi )}{\pi\Delta L - \sin\theta}, \text{ for }  |\psi^-\rangle. \nonumber
\end{align}
As illustrated in Fig.~\ref{fig:fig1}, this expression violates Bell inequalities for any finite $\Delta L$. While it cannot reproduce the results of quantum mechanics exactly if we assume independence of the $\Delta L$ between detectors (solving for $\Delta L$ in $E^{hv}_L(\theta) = - \cos\theta$ yields a non-separable function of $\theta$), Eq.~\eqref{eq:ehvl} can closely match $E^{qm}(\theta)$ with a constant $\Delta L$ that is assumed to be small; $\Delta L \approx 0.77$ gives a mean error of $\approx 0$ with a maximum disagreement of $\approx 0.03$. Lower values of $\Delta L$ allow for violations of Tsirelson's bound~\cite{Tsirelson1980}. The minimum physically meaningful value of $\Delta L = 0.5$ prevents $|E^{hv}_L(\theta)| > 1$  and can be associated with a measurement where a single qubit $|\alpha^\pm\rangle$ acts as the probe.

The requirement of a small $\Delta L$ contrasts with the usual discussions of the WAY theorem~\cite{Wigner1952,Araki1960,Marvian2012,Loveridge2011}, which assume the probe to be of macroscopic scale, and $\Delta L \to \infty$ (and thus an accurate measurement). Nonetheless, the boundary defining the probe has always been rather arbitrary~\cite{Ozawa2002}. A $\Delta L$ of the magnitude required here for observing strong violations of Bell inequalities would require a new model for the interaction that determines the measurement outcome, one that is extremely local and on a single or few particles level for the probe. 

For the Bell singlet, all the contributions to $\Delta L$ come from the probe. This leads to comparatively weaker correlations predicted for the Bell triplet states $|\psi^+\rangle = (|01\rangle + |10\rangle)/\sqrt{2}$ and $|\phi^-\rangle = (|00\rangle - |11\rangle)/\sqrt{2}$ that have $\Delta L = \Delta J_y = 1$. We can perform an analysis analogous to the one for $|\psi^-\rangle$ for these two states with the same deviation operator $D$ as in Eq.~\eqref{eq:deviationD}, to obtain
\begin{equation}
  E^{hv}_L(\theta) = \pm\frac{(\Delta L_\xi + 1) (2\theta - \pi )}{\pi\Delta L_\xi  - \sqrt{3}\sin\theta + \pi}, \text{ for }  |\psi^+\rangle  \text{ and } |\phi^-\rangle. 
\end{equation}
This expression is also illustrated in Fig.~\ref{fig:fig1} and compared to that in Eq.~\eqref{eq:ehvl}. Assuming that $\Delta L_\xi$ is independent of the state in which the system being measured was initially prepared, the differences predicted between singlet and triplet states could be detected by sufficiently sensitive instruments with high-fidelity state preparations (making the model an extension rather than an interpretation of quantum mechanics, in the language of \cite{adlam2023taxonomy}). Similarly, as the size of the system being measured increases, we expect diminished correlations that approach the classical limit of Eq.~\eqref{eq:EHV} as $\Delta L \to \infty$.  

Of course, single-spin measurements would also need to be explained by the hidden-variable theory. As noted by Bell~\cite{Bell1964}, there is no difficulty in describing the results of single spin measurements with hidden variables; contradictions only arise when considering either multi-qubit entangled systems, or systems of dimension $\geq3$ \cite{budroni2022contextualityReview}. We could use the same WAY-based arguments used above to explain the results of measuring a single spin. Consider a qubit in the initial state $|0\rangle$ and a measurement of $\sigma_\alpha$. A deviation operator analogous to that of Eq.~\eqref{eq:deviationD} can be defined as $D = U_\alpha^\dag(\sigma_z \otimes I) U_\alpha - \sigma_\alpha \otimes I$, where $U_\alpha$ takes $|0\rangle$ and the probe to the post-measurement state. One can then follow the same procedure as above for the singlet Bell state and get the expectation values of quantum mechanics. The only difference is that the angle between detectors $\theta$ is replaced by $\alpha$, which corresponds to the difference in the spin orientation of the initial and final states. Contrary to the two-body experiment, there are no restrictions on the dependence of $\Delta L_\xi$ on $\alpha$ in the single-particle case so that the results of quantum mechanics can be reproduced exactly. 

\emph{Discussion---}As we alluded earlier, the argument given here for the possibility of hidden variables is closely related to the well-known detection loophole. Not long after Bell's landmark paper, Pearle~\cite{Pearle1970} formulated a hidden-variables model that reproduced the results of quantum mechanics by discarding certain experimental outcomes. High-efficiency detectors are considered to have closed this loophole~\cite{Rowe2001,Giustina2015,Hensen2015}. However, as suggested here, it is possible that a specific range of outcomes cannot be observed (not even in principle) because they do not preserve global symmetries during the measurement process. Which outcomes are unseen is determined by the global measurement settings, which dictate the symmetry of the system combined with the measurement probe. This alters the measure space $(\Lambda, \mathcal{A}, \mu)$, but the underlying probability density function $\rho(\lambda)$ maintains statistical independence from the detector settings~\cite{Hance2022}. Therefore, the model is what has been referred to in the literature as ``supermeasured''---violating the Statistical Independence assumption required to generate Bell inequalities, but doing so for a different physical reason than the two most famous classes of model which do this (superdeterminism and retrocausality).

We'd like to emphasize that this model is Bell-nonlocal, but only as it violates (Bell-)Statistical Independence for the physically well-motivated reason of preserving both conservation laws (and so symmetry) and the uncertainty principle simultaneously. While other Toy Models exist, to show other ways that local hidden variable models can violate Bell inequalities through violating Statistical Independence (e.g., \cite{Brans1988Model,Palmer1995Spin,Degorre2005Sampling,Palmer2009ISP,Hall2010Model,Donadi2020SuperdetToy,Ciepielewski2020Superdeterministic}), these tend to be physically ad-hoc. However, our model is based on physical principles which explain why we would need to make a local hidden variable model supermeasured.

If there were no relevant global symmetries to be conserved, then the argument we have employed here for modifying the measure space $(\Lambda, \mathcal{A}, \mu)$ would not hold. Recent loophole-free experiments with superconducting qubits~\cite{Storz2023} have shown weak violations of Bell's inequality  ($S$ value of 2.0747 in the  Clauser--Horne--Shimony--Holt~\cite{CHSH1969} inequality). The conservation of global angular momentum that we have here considered is not relevant to superconducting qubits. However, the WAY theorem is relevant for other conserved quantities such as energy~\cite{Katsube2022} and momentum~\cite{Kuramochi2023}, and it can be generalized in terms of symmetry and quantum incompatibility~\cite{Tukiainen2017}. More generally, it is well-known in quantum mechanics that no information can be gained without disturbance of the measured system~\cite{Busch2009}. Restrictions on the amount of information that can be extracted from a system and the uncertainty principle must be respected in some way even in hidden-variable theories, as otherwise one could devise thermodynamic cycles that violate the second law of thermodynamics~\cite{Hanggi2013}. Thus, while we do not suggest that the formulation given here of a local hidden-variable theory that can violate Bell inequalities due to restrictions from conservation laws is necessarily a correct description of physical reality (that would indeed seem rather unlikely), further studies may be needed to completely discard local realism. Of particular relevance would be the study of how symmetry preservation may affect measurement and the measure space of the hidden variables, as well as re-analyzing data from existing experiments or devising new experiments that could test predictions such as the subtle differences in correlations between Bell singlet $|\psi^-\rangle$ and triplet $|\psi^+\rangle$ states predicted here.

\emph{Acknowledgments---}JRH acknowledges support from a Royal Society Research Grant (RG/R1/251590), and from their EPSRC Quantum Technologies Career Acceleration Fellowship (UKRI1217).

\emph{Code Availability Statement---}All code generated during this work is available from the authors on request.

\bibliographystyle{unsrturl}
\bibliography{ref.bib}

\end{document}